\documentclass[aps,prd,twocolumn,superscriptaddress,showpacs,nofootinbib]{revtex4-1}

\usepackage[utf8]{inputenc}

\usepackage{diagbox}

\usepackage{mathtools}
\usepackage{amsfonts}
\usepackage{mathrsfs}
\usepackage{bbm}
\usepackage{slashed}

\usepackage{graphicx}
\usepackage{color}
\usepackage{array}
\usepackage{esint}
\usepackage{placeins}
\usepackage{booktabs}
\usepackage{makecell}
\usepackage{epstopdf}
\usepackage[caption=false]{subfig}

\usepackage{xspace}
\usepackage{siunitx}
\usepackage{hyperref}
\usepackage[nameinlink]{cleveref}
\usepackage{appendix}

\usepackage{xifthen}
\usepackage{xcolor}
\hypersetup{
	colorlinks,
	linkcolor={red!75!black},
	citecolor={blue!75!black},
	urlcolor={blue!75!black}
}

\setkeys{Gin}{width=0.48\textwidth}

\graphicspath{
	{./figures/}
	{../figures/}
}

\def\s0#1#2{\mbox{\small{$ \frac{#1}{#2} $}}}
\def\0#1#2{\frac{#1}{#2}}

\def\Eq#1{Eq.~\eqref{#1}}

\def\Fig#1{Figure~\ref{#1}}

\def\Sec#1{Section~\ref{#1}}
\newcommand{\gettitle}{A correspondence between the free and interacting field theories}

\hypersetup{
	pdftitle={\gettitle},
	pdfauthor={Gao, Pawlowski},
	pdfkeywords={functional methods}
		{correlations functions}  {liquid gas transition}
		{functional renormalisation group}{Dyson-Schwinger equations},
	bookmarksopen=true,
	bookmarksopenlevel=2,
	bookmarksnumbered=true
}

\begin{document}

\title{\gettitle}

\author{Fei Gao}
\email{hiei@pku.edu.cn}
\affiliation{Center for High Energy Physics, Peking University,
	100871 Beijing, China
}

\author{Minghui Ding}
\email{m.ding@hzdr.de}
\affiliation{Helmholtz-Zentrum Dresden-Rossendorf, Bautzner Landstra{\ss}e 400,
01328 Dresden, Germany
}

\author{Yu-xin Liu}
\email{yxliu@pku.edu.cn}
\affiliation{Center for High Energy Physics, Peking University,
	100871 Beijing, China
}
\affiliation{Department of Physics and State Key Laboratory of Nuclear Physics and Technology, Peking University, Beijing 100871, China}
\affiliation{Collaborative Innovation Center of Quantum Matter, Beijing 100871, China}

\author{Sebastian \,M.~Schmidt}
\email{s.schmidt@hzdr.de}
\affiliation{Helmholtz-Zentrum Dresden-Rossendorf, Bautzner Landstra{\ss}e 400,
01328 Dresden, Germany
}
\affiliation{RWTH Aachen University, III. Physikalisches Institut B, Aachen D-52074, Germany
}
%\pacs{Valid PACS appear here}% PACS, the Physics and Astronomy
%\pacs{12.38.Aw, %General properties of QCD (dynamics, confinement, etc.)}                             % Classification Scheme.
%\keywords{Suggested keywords}%Use showkeys class option if keyword
%display desired

\date{\today}

\begin{abstract}
We discover a correspondence between the free field and the interacting states. This correspondence is firstly given from the fact  that the free propagator can be converted into a tower of propagators for massive states, when expanded with the Hermite function basis. The equivalence of propagators reveals that  in this particular case the duality can naturally be regarded as the equivalence of one theory on the plane wave basis to the other on the Hermite function basis. More generally, the Hermite function basis provides an alternative quantization process with the creation/annihilation operators  that correspond directly to the interacting fields.  Moreover, the Hermite function basis defines an exact way of dimensional reduction. As an illustration, we apply this basis on 3+1 dimensional Yang-Mills theory with  three dimensional space being reduced through the Hermite function basis, and  if  with  only the lowest order Hermite function, the equivalent action becomes the Banks-Fischler-Shenker-Susskind (BFSS) matrix model.
\end{abstract}

\maketitle

\section{Introduction}\label{sec:Introduction}
There have been longstanding efforts for  understanding the phenomenon of duality. Generally speaking, duality is a way of showing  the correspondence between two apparently different theories.  The aspects covered by duality are quite numerous and include the target space  duality~\cite{Giveon:1994fu,Alvarez:1994dn,Harvey:1995tg},  strong-weak  duality~\cite{Montonen:1977sn,Sen:1994fa,Alvarez:1989ad} and fermion-boson duality~\cite{Coleman:1974bu,Mandelstam:1975hb,Fradkin:1994tt,Aharony:2015mjs,Karch:2016sxi}, and so on.

Especially, a large class of duality has been known as particle-vortex duality, which  is the dual of the Higgs model with the XY model~\cite{Dasgupta:1981zz,Peskin:1977kp} for bosonic system, and the dual of the Dirac fermion with the composite one~\cite{Mross:2015idy,Son:2015xqa,Wang:2015qmt,Cheng:2016pdn,Senthil:2018cru} for the fermionic case.
It was later realised that this type of duality could generally be grouped into the fermion-boson duality by bosonisation~\cite{Karch:2016sxi,Seiberg:2016gmd}. The idea was to attach the flux to fields that had been found to switch the statistical transmutation of particles~\cite{Wilczek:1981du,Polyakov:1988md,Shaji:1990is,Paul:1990vw}. After attaching the flux, the new state defined by the monopole operator which carries a transmutation different from that of the fundamental field emerges.
The underlying concept of these dualities is  the relation between the fundamental field and the interacting field associated with the flux.

  Inspired by this,  we propose a new approach based on Hermite function basis, and  discover a closely related correspondence between the free field and the interacting states. This correspondence shows the equivalence of  two theories, one of which can be achieved by expanding the other on the Hermite function basis. The Hermite functions are the eigenfunctions of the harmonic oscillator in quantum mechanics,  forming a complete orthonormal basis. The formulae of harmonic oscillators or Hermite functions  is ubiquitous and can be   intuitively   interpreted as  the basis of interacting field. Based on these observations, one may  expand the free field on account of the Hermite function basis being orthonormal, and then   the resulting theory is supposedly  converted into the interacting field picture.

It becomes clear in the canonical quantization procedure,  whereas the field holds the same canonical commutation relation, on the Hermite function basis, its respective  creation/annihilation operators  now directly describe the harmonic type interacting particles.  By applying this, we study the correspondence for both fermion and boson fields. We hope this will shed some light on the understanding of the phenomenon of duality.

At last, we illustrate the application of this approach on the 3+1 dimensional Yang-Mills theory. After reducing the three dimensional space with the Hermite function basis and constraining to the lowest order Hermite function, the resulting action becomes the action of the Banks-Fischler-Shenker-Susskind (BFSS) Matrix model derived through dimensional reduction ~\cite{Banks:1996vh,Bilal:1997fy,Connes:1997cr,Bergner:2019rca}. Naturally, this approach  offers an exact way of dimensional reduction without requiring compactification of the space~\cite{Klein:1926tv,Freund:1980xh,Witten:1985xb}.

 The article is organised as follows. In \Sec{sec:Derivations} we start with the equation of motion for fermion/boson, and derive the respective propagator on the Hermite function basis. Then, in \Sec{sec:canonical} we lead up to an alternative quantization process and observe a duality relation in the action of the field. In \Sec{sec:YM} we illustrate the correspondence between the 3+1 dimensional Yang-Mills theory and the one dimensional Matrix model. In \Sec{sec:summary} we summarise our approach and present our conclusions.

\section{Equation of motion in the Hermite function basis}\label{sec:Derivations}

We start our discussion with the equation of motion for fermions, i.e. the Dyson-Schwinger equation (DSE) for the fermion propagator.   The DSE for the propagator of the fermion interacting with the gauge field is generally written as:
\begin{eqnarray}
\label{eq:dse}
&&(-\partial_a \gamma_z +i\bar{p}\!\!\!/+m_0)S(\bar{p};a-b)=\delta(a-b)+g^2\int d k_zd^3\bar{q}d c \notag\\
&&\times\gamma_\mu S(\bar{q};a-c)\Gamma_\nu D^{\mu\nu}(\bar{p}-\bar{q},k_z)e^{-ik_z(a-b)}S(\bar{p};c-b)\,,
\end{eqnarray}
with $S$ the fermion propagator; $D^{\mu\nu}$ the gauge boson propagator; $\Gamma_\mu$ the full interaction vertex with tree level as $\gamma_\mu$; $g$ the running coupling; $\delta$ the Dirac delta function and the metric being set as $(-1,1,1,1)$. Here the DSE is in a mixed representation with momentum representation for $\bar{p}=(p_t,p_x,p_y,0)$ and coordinate representation for $z-$axis ($a$ and $b$ are on the $z-$axis in the coordinate space). The integrated momentum are $\bar{q}=(q_t,q_x,q_y,0)$, and $k_z=p_z-q_z$, where $p_z,\,q_z$ and $k_z$ are on the $z-$axis in the momentum space. If the Fourier transform is also applied to the $z-$axis, the usual DSE in momentum representation can be obtained~\cite{Roberts:1994dr,Alkofer:2000wg,Eichmann:2016yit, Binosi:2014aea,Williams:2015cvx,Aguilar:2018epe,Tang:2019zbk,Qin:2020jig,Gao:2021wun,Chang:2021vvx}. Here we are taking the case where the $z$-direction is in coordinate space as an example, and the approach is applicable and can be extended to other cases where other directions ($t, x, y$) are in coordinate space.

The equation for the free fermion propagator of zero coupling ($g=0$) is then given by:
\begin{eqnarray}\label{eq:free}
(-\partial_a \gamma_z +i\bar{p}\!\!\!/+m_0)S(\bar{p};a-b)=\delta(a-b)\,.
\end{eqnarray}
With the momentum representation, the free fermion propagator can be obtained as:
\begin{eqnarray}\label{eq:freemom}
S(\bar{p};a-b)=\int dp_z\frac{ e^{-ip_z(a-b)}}{ip\!\!\!/_z+i\bar{p}\!\!\!/+m_0}\,,
\end{eqnarray}
 which may be  formally  regarded as being expanded with the plane wave basis. Here instead, as we  described in preceding section, the Hermite functions also form a complete orthonormal basis, and we therefore use the Hermite function basis to expand the fermion propagator on the $z-$axis, i.e.,
\begin{eqnarray}
\label{eq:ritus}
&&S(\bar{p};a-b)=\sum_n S_n(\bar{p})f_n(a-b)\notag\\
=&&\sum_n \left[-i \sigma^{A}_{n}(\bar{p}^2)\bar{p}\!\!\!/+\sigma^{B}_{n}(\bar{p}^2)-i\sigma^{C}_{n}(\bar{p}^2)\gamma_z\right]f_n(a-b),
\end{eqnarray}
where the general form of $\bar{p}-$dependent scalar functions are denoted by $\sigma^{A}_{n}(\bar{p}^2), \sigma^{B}_{n}(\bar{p}^2)$, and $\sigma^{C}_{n}(\bar{p}^2)$ respectively; $f_n(z)$ with $n\geq0$ is the  Hermite function (sometimes called Hermite-Gaussian function) as:
\begin{equation}\label{eq:Hermite}
f_n(z)=\frac{\omega^{1/2}}{\sqrt{2^n n!\sqrt{\pi}}}H_n({\omega}z)e^{-\omega^2 z^2/2},
\end{equation}
and $H_n(z)$ is the Hermite polynomial and the orthogonal normalisation condition is:
\begin{equation}
\label{eq:prop1}
\int dz f_n(z)f_m(z)=\delta_{nm}\,.
\end{equation}
The Hermite function, which gives rise to the wave function of the energy eigenstate of the quantum harmonic oscillator, when operated by the creation and annihilation operators ($a^\dagger=\frac{\omega^2 z-\partial_z}{\sqrt{2}{\omega}}, a=\frac{\omega^2 z+\partial_z}{\sqrt{2}{\omega}}$), give:
\begin{eqnarray}
\label{eq:prop2}
&&\frac{\omega^2 z-\partial_z}{\sqrt{2}{\omega}}f_n(z)=\sqrt{n+1}f_{n+1}(z)\,,\notag\\
&&\frac{\omega^2 z+\partial_z}{\sqrt{2}{\omega}}f_n(z)=\sqrt{n}f_{n-1}(z)\,.
\end{eqnarray}
It should be mentioned that in \Eq{eq:ritus} there exists in principle $\gamma_z \bar{p}\!\!\!/$ term, but this can be excluded by comparing it with the momentum representation of the fermion propagator in \Eq{eq:freemom}.

By now we have expanded the fermion propagator with the Hermite function basis on the $z-$axis in the coordinate space. Actually this Hermite function basis expansion method has been broadly used in the studies of continuum Schwinger method with a constant background magnetic field, namely the  Ritus formula~\cite{Ritus:1972ky,
Mueller:2014tea,Xing:2021kbw}.  In a constant magnetic field, the Hermite function basis can be closed to four terms, including only $f_n$, $f_{n-1}$, and $f_{n+1}$. Here we generalize this method by inserting an auxiliary field $a$ and  decomposing the operator on the left hand side of \Eq{eq:free} as:
\begin{eqnarray}
\label{eq:decom}
&&-\partial_a \gamma_z +i\bar{p}\!\!\!/+m_0\notag\\
=&&
-\frac{1}{2}(\partial_{a}+\omega^2 a)\gamma_z-\frac{1}{2}(\partial_{a}-\omega^2 a)\gamma_z+i\bar{p}\!\!\!/+m_0.
\end{eqnarray}
This decomposition involves only the differential operator and does not require the  background field. The Hermite function basis cannot now be closed within finite terms, thus it requires to use infinite dimensional matrices defined as linear operators for the functional analysis. To make this point clearer, one may first insert the fermion propagator expansion expression in \Eq{eq:ritus} into the equation it satisfies, i.e., DSE in \Eq{eq:free}, and apply the property  of Hermite function in \Eq{eq:prop2} and the decomposition in \Eq{eq:decom} to obtain:
 \begin{widetext}
 \begin{eqnarray}
\label{eq:mdse0}
&&\sum_{m^\prime}\big{\{} (i\bar{p}\!\!\!/+m_0)\left[-i \sigma^A_{m^\prime}(\bar{p}^2)\bar{p}\!\!\!/+\sigma^B_{m^\prime}(\bar{p}^2)-i\sigma^C_{m^\prime}(\bar{p}^2)\gamma_z\right]f_{m^\prime}(a-b)-\frac{\gamma_z}{2}\sqrt{2m^\prime}\omega\left[-i \sigma^A_{m^\prime}(\bar{p}^2)\bar{p}\!\!\!/+\sigma^B_{m^\prime}(\bar{p}^2)-i\sigma^C_{m^\prime}(\bar{p}^2)\gamma_z\right]\notag\\
&&\times f_{m^\prime-1}(a-b)+\frac{\gamma_z}{2}\sqrt{2(m^\prime+1)}\omega\left[-i \sigma^A_{m^\prime}(\bar{p}^2)\bar{p}\!\!\!/+\sigma^B_{m^\prime}(\bar{p}^2)-i\sigma^C_{m^\prime}(\bar{p}^2)\gamma_z\right]f_{m^\prime+1}(a-b)\big{\}}=\delta(a-b)\,.
\end{eqnarray}
 \end{widetext}

 Multiplying the resulting \Eq{eq:mdse0} by the integral $\int  d a f_m(a-b)$, and applying the orthogonal normalisation condition in \Eq{eq:prop1}, one get that all the Hermite functions turn to the Dirac delta functions, which may be combined with the index $m^\prime$ in the scalar functions $\sigma^{A,B,C}_{m^\prime}(\bar{p}^2)$.
%\begin{eqnarray}
%\label{eq:mdse}
%&&\sum_{m^\prime}\big{\{} (i\bar{p}\!\!\!/+m_0)(-i \sigma^A_{m^\prime}(\bar{p}^2)\bar{p}\!\!\!/+\sigma^B_{m^\prime}(\bar{p}^2)-i\sigma^C_{m^\prime}(\bar{p}^2)\gamma_z)\delta^{m,m^\prime}\notag\\
%&&-\frac{\gamma_z}{2}\sqrt{2m^\prime}\omega(-i \sigma^A_{m^\prime}(\bar{p}^2)\bar{p}\!\!\!/+\sigma^B_{m^\prime}(\bar{p}^2)-i\sigma^C_{m^\prime}(\bar{p}^2)\gamma_z)\delta^{m,m^\prime-1}\notag\\
%&&+\frac{\gamma_z}{2}\sqrt{2(m^\prime+1)}\omega\notag\\
%&&\times(-i \sigma^A_{m^\prime}(\bar{p}^2)\bar{p}\!\!\!/+\sigma^B_{m^\prime}(\bar{p}^2)-i\sigma^C_{m^\prime}(\bar{p}^2)\gamma_z)\delta^{m,m^\prime+1}\big{\}}\notag\\
%=&&f_m(0)
%\end{eqnarray}
%-g^2C_F\int d^3\bar{q}\sum_{nn^\prime}\big{\{}\gamma_\mu(-i \sigma^{A}_{n}(\bar{q}^2)\bar{q}\!\!\!/+\sigma^{B}_{n}(\bar{q}^2)-i\sigma^{C}_{n}(\bar{q}^2)\gamma_z)\notag\\
%&&\times\mathcal{I}_\mu^{nn^\prime m}(\bar{p}-\bar{q})(-i \sigma^{A}_{n^\prime}(\bar{p}^2)\bar{p}\!\!\!/+\sigma^{B}_{n^\prime}(\bar{p}^2)-i\sigma^{C}_{n^\prime}(\bar{p}^2)\gamma_z)\big{\}}
%with $$\mathcal{I}^{nn^\prime m}_{\mu}(\bar{p},\bar{q})=\int dk_4 da dc \Gamma_\nu(\bar{p},\bar{q},k_4) $$ $$D_{\mu\nu}(\bar{p}-\bar{q},k_4)e^{ik_4(a-c)}f_n(a-c)f_{n^\prime}(c-b)f_m(a-b)$$.
Then  comparing the corresponding Dirac terms on both sides of the equation, one can directly obtain (We drop the explicit $\bar{p}$ index in the scalar functions $\sigma^{A,B,C}_{m}$ and it is included implicitly):
\begin{subequations}
\begin{eqnarray}
\label{eq:abc}
f_m(0)=&&\,\bar{p}^2\sigma^A_m+m_0\sigma^B_m+i\sqrt{\frac{m+1}{2}}\omega\sigma^C_{m+1}\notag\\
&&\,-i\sqrt{\frac{m}{2}}\omega\sigma^C_{m-1}\,,\\
0= &&\,m_0\sigma^A_m-\sigma^B_m\,,\\
0= &&\,-i\sqrt{\frac{m+1}{2}}\omega\sigma^A_{m+1}+i\sqrt{\frac{m}{2}}\omega\sigma^A_{m-1}+\sigma^C_m\,.
\end{eqnarray}
 \end{subequations}

Let's now define an operator $\hat{T}$  as:
\begin{equation}
\hat{T}=T_{mm^\prime}=i\sqrt{\frac{m^\prime}{2}}\omega\delta^{m,m^\prime-1}-i\sqrt{\frac{m}{2}}\omega\delta^{m,m^\prime+1},
\end{equation}
which acts on the Hermite function basis space. One can  represent it  in  the infinite dimensional matrix form as:
\begin{equation}
\label{eq:matrix}
\begin{gathered}
\begin{bmatrix}
0 &  \frac{i\omega}{\sqrt{2}} & 0 & 0& 0 & 0  \\
-\frac{i\omega}{\sqrt{2}}&0 & i \omega & 0 & 0& 0 &  \\
0&-i \omega & 0 &i\sqrt{\frac{3}{2}}\omega& 0& 0 & \\
0&0 & -i\sqrt{\frac{3}{2}}\omega &0& i\sqrt{2}\omega& 0 &  \\
0 &0& 0&-i\sqrt{2}\omega &0& i\sqrt{\frac{5}{2}}\omega&   \\
0 &0& 0&0&-i\sqrt{\frac{5}{2}}\omega& 0&\cdots \\
 && &  & & \cdot &  \\
 && &   && \cdot & \\
\end{bmatrix}
\end{gathered}
\end{equation}
The free fermion propagator can then be conveniently expressed as:
\begin{equation}
\label{eq:hfermion}
S(\bar{p}; a-b)=\vec{f}(a-b)\frac{-i \hat{T}\gamma_z-i\bar{p}\!\!\!/+m_0}{\bar{p}^2+m^2_0+\hat{T}^2}\vec{f}(0),
\end{equation}
with $\vec{f}=\{f_0,f_1,...f_n\}$ the array of Hermite functions in \Eq{eq:Hermite}.  It follows from \Eq{eq:matrix} that $\hat{T}$ is a Hermitian operator, so that it can be diagonalised as $\hat{T}=\hat{P}^\dagger \Omega_n \hat{P}$. Interestingly, for even basis, $\hat{T}$ has two sets of eigenvalues that differ only by the sign $\pm1$, i.e., $\Omega_n=\pm|\Omega_n|$. For odd basis, there is an additional eigenvalue of zero.

\begin{figure*}[t]
\includegraphics[width=0.95\columnwidth]{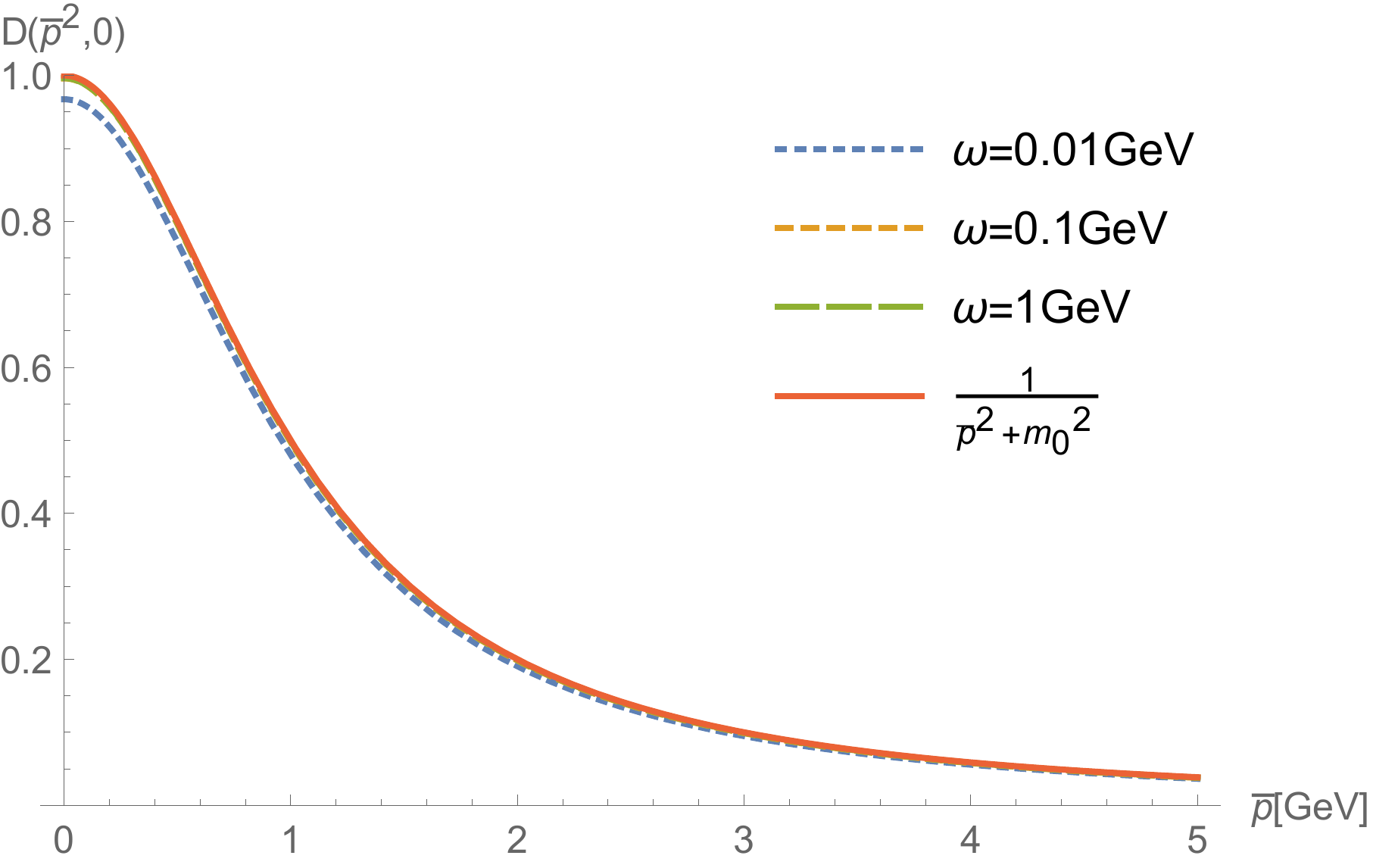}
\includegraphics[width=0.95\columnwidth]{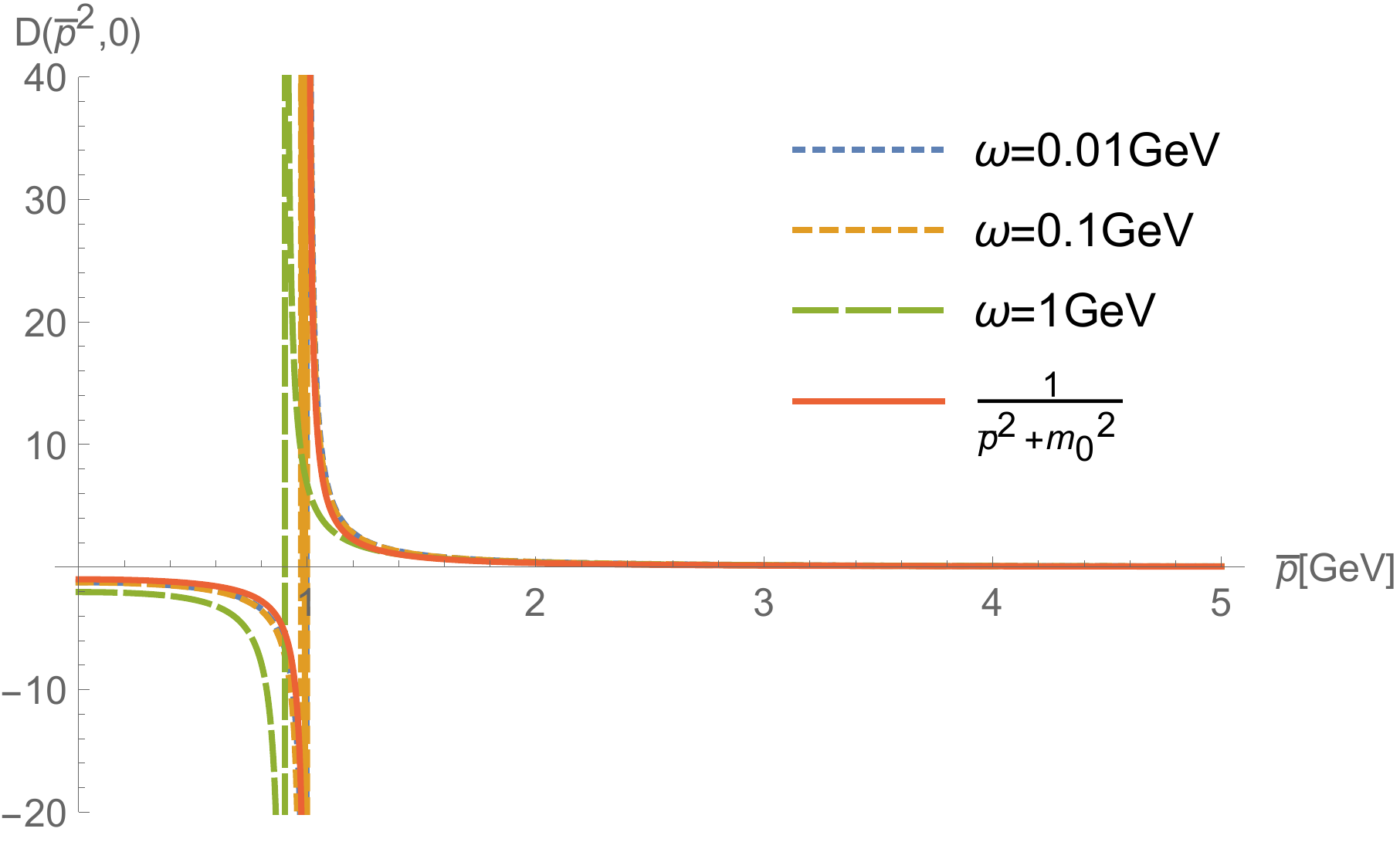}
	\caption{The boson propagator obtained from the Hermite function basis expansion as in  \Eq{eq:propcor} compared to the free boson propagator $1/(\bar{p}^2+m^2_0)$ expanded on the plane wave basis, with two different types of masses, real mass with  $m_0=1$ GeV (\emph{left panel}) and purely imaginary mass with $m_0=i$ GeV (\emph{right panel}).  }
	\label{fig:corr}
\end{figure*}

 In order to build a relation with the momentum representation of the free fermion propagator, as in \Eq{eq:freemom}, it is convenient to take the Fourier transform of $a-b$ in \Eq{eq:hfermion}, i.e. a transformation of the $z-$axis in coordinate space. The Fourier transform of the Hermite function here is
 \begin{eqnarray}
 \int da e^{-ip_z(a-b)}f_n(a-b)=\frac{\sqrt{2\pi}(-i)^n}{\omega} f_{n}\left(\frac{p_z}{\omega^2}\right)\,,
\end{eqnarray}
 and note that the odd Hermite function vanishes at the origin, and therefore  for $p_z=0$  we only need to consider the even Hermite function. We denote $\tilde{\Omega}_n=\Omega_{2n}$ for convenience and obtain:
\begin{eqnarray}
\label{eq:propcorfer}
&&S(\bar{p}, p_z=0)=\frac{-i\bar{p}\!\!\!/+m_0}{\bar{p}^2+m^2_0}\notag\\
=&&(-i\bar{p}\!\!\!/+m_0)\frac{\sqrt{2\pi}}{\omega}\sum_n\frac{\hat{\sigma}(\varphi^n)^*(\varphi^n)}{\bar{p}^2+m^2_0+\tilde{\Omega}^2_{n}}\,,
\end{eqnarray}
with
\begin{eqnarray}
\varphi^n=\hat{P} f_{2n}\left(\frac{p_z}{\omega^2}=0\right)\,.
\end{eqnarray}
The operator $\hat{\sigma}$ acting on the $4n$ and $4n+2$ basis function will give $1$ and $-1$ respectively.   Leaving aside the Lorentz structure $-i\bar{p}\!\!\!/+m_0$, the propagator corresponds to that of the interacting theory, describing a tower of massive states of mass spectrum $\tilde{\Omega}_n$ and  wave function $\varphi^n$  with norm operator $\hat{\sigma}=\pm1$. In a sense, it corresponds to two towers of massive states with positive and negative norms, that ultimately cancel and maintain a free propagation mode. Here we use the case $p_z=0$ as an example, the discussion is applicable and can be extended to other cases where $p_z$ is non-zero.

Our work from the beginning of this section onwards has all been concerned with the fermion field. It gave a general correspondence in \Eq{eq:propcorfer} revealing that the fermion propagator of the free field corresponds to that of the interacting field of massive states. To get a complete image we then consider also the correspondence for the bosonic propagators.

Again, we start from the equation for the free boson propagator $D$ of zero coupling given by:
\begin{eqnarray}\label{eq:free1}
(-\partial^2_a+\bar{p}\!\!\!/^2+m^2_0)D(\bar{p};a-b)=\delta(a-b)\,.
\end{eqnarray}
Similarly, we expand the boson propagator using the Hermite function basis as:
\begin{eqnarray}
D(\bar{p};a-b)=\sum_n D_n(\bar{p})f_n(a-b),
\end{eqnarray}
and performing the same procedure as in the fermion case above, we obtain the boson propagator, expressed as:
\begin{eqnarray}
D(\bar{p};a-b)=\vec{f}(a-b)\frac{1}{\bar{p}^2+m^2_0+\hat{T}^2}\vec{f}(0).
\end{eqnarray}
Comparing it with the momentum representation of the free boson propagator, we obtain:
%\begin{widetext}
\begin{eqnarray}
\label{eq:propcor}
&&D(\bar{p},p_z=0)=\frac{1}{\bar{p}^2+m^2_0}=\frac{\sqrt{2\pi}}{\omega}\sum_n\frac{\hat{\sigma}(\varphi^n)^*(\varphi^n)}{\bar{p}^2+m^2_0+\tilde{\Omega}^2_{n}}\,.\notag\\
\end{eqnarray}
%\end{widetext}
This gives an analogous correspondence between the boson propagator of the free field and the interacting field with a tower of massive states. Thus we note that a correspondence also exists in the boson system.

 It's worth mentioning that the parameter $\omega$ in the Hermite function basis representation is in fact arbitrary, i.e., \Eq{eq:propcor} can be satisfied for any $\omega$ if the infinite order of the Hermite function is used. This is rather nontrivial, since $\omega$ in \Eq{eq:propcor} is not easy to be fully cancelled. Therefore, we try to verify the correspondence in \Eq{eq:propcor} numerically  with two very different types of propagators, the results of which are shown in \Fig{fig:corr}. In the left panel of \Fig{fig:corr}, the original propagator is like a propagator in Euclidean momentum space, with $m_0=1$ GeV, while in the right panel it has $m_0=i$ GeV, like a propagator in Minkowski space, with a pole on the momentum axis.
To begin with, we find that the correspondence is indeed generally independent of the choice of $\omega$. We considered the correspondence under three different $\omega$, namely $\omega=0.01, 0.1, 1$ GeV, for all of which \Eq{eq:propcor} is approximately satisfied. There are some deviations due to the truncation of the basis space of Hermite functions. To be specific, for the real mass case, a smaller $\omega$ brings more deviation, while for the purely imaginary mass, a smaller $\omega$ implies a better correspondence. Additionally, it is interesting to note that the space required for the Hermite function is different in these two cases. Specifically, for $m_0=1$ GeV, we apply here the Hermite function $f_n(z)$ of order up to $n=100$, while for the propagator with a pole on the momentum axis, i.e. $m_0=i$ GeV, a good correspondence can be achieved with very few orders of the Hermite function for sufficiently small $\omega$. For example, at $\omega=0.01$ GeV, we apply the first six orders of the Hermite function as $f_{0,1,\cdots 5}(z)$. This indicates that the Hermite function basis might be useful for nonperturbative studies towards Minkowski space, such as the spectrum of states or the transport properties of the system.

\section{Canonical quantization on Hermite function}\label{sec:canonical}
We further investigate the Hermite function basis more generally by the action. Clearly, an alternative typical canonical quantization based on the Hermite function basis can be constructed. Let's begin with the free fermion field. The basic idea is to expand the $d+1$ dimensional free fermion field $\psi$ as:
\begin{eqnarray}
\label{eq:expand}
&&\psi\left(t,x_1,\cdots x_p,x_{p+1} \cdots x_{d}\right)
=\sum_{n_1,\cdots n_p} \notag\\
&& \Psi_{n_1,\cdots n_p}\left(t,x_{p+1},\cdots x_{d}\right)[\hat{P}f]_{n_1}(x_1)\cdots[\hat{P}f]_{n_p}(x_p).
\end{eqnarray}
$\hat{P}$, as described above, is the matrix acting on the Hermite function basis space to diagonalize the operator $\hat{T}$ defined in \Eq{eq:matrix}. $\Psi$ is a new field, the part that has not been expanded. Here we expand the fermion field in the direction $n_1$ to $n_p$, this is used as an example, in fact it can be extended to the expansion in other directions as well.  The canonical anti-commutation relation of the original free fermion field $\psi$ is:
\begin{eqnarray}
\label{eq:anticomm}
\left\{\psi(t,x_1,\cdots x_{d}),\psi^\dagger(t,y_1,\cdots y_{d})\right\}=\delta^d(x-y).
\end{eqnarray}
We can set the  new field $\Psi$ to meet the following constraint:
\begin{eqnarray}
\label{eq:quanta}
&&\left\{  \Psi_{n_1,\cdots n_p}(t,x_{p+1},\cdots x_{d}),\Psi^\dagger_{m_1,\cdots m_p}(t,y_{p+1},\cdots y_{d})\right\}\notag\\
=&&\delta^{n_1,m_1}\cdots \delta^{n_p,m_p}\delta(x_{p+1}-y_{p+1})\cdots\delta(x_d-y_d),
\end{eqnarray}
and then use the properties of the Hermite function $\sum_n f_n(x)f_n(y)=\delta(x-y)$ and the matrix used for diagonalisation $\hat{P}\hat{P}^\dagger=\mathcal{I}$, the anti-commutation relation in \Eq{eq:anticomm} can be reproduced. Comparing \Eq{eq:quanta} with \Eq{eq:anticomm}, we see that the  new field holds a similar canonical anti-commutation relation as that of the original free fermion field, hence they must be connected in some way, at least this suggests that the new field is also a fermion field.

Let's now consider the action of the $d+1$ dimensional free fermion field:
\begin{eqnarray}
\label{eq:fermionaction}
\mathcal{S}=\int dt d^{d}x\left[-\bar{\psi}(x)\partial\!\!\!/\psi(x)+m_0\bar{\psi}(x)\psi(x)\right]\,.
\end{eqnarray}
Substituting \Eq{eq:expand} in \Eq{eq:fermionaction}, the action is converted to:
\begin{eqnarray}
\label{eq:fermionactionnew}
\mathcal{S}=&&\int dt d^{d-p} x\sum_n\big[-\bar{\Psi}_n(x)\partial\!\!\!/\Psi_n(x)\notag\\
&&-i\bar{\Psi}_n(x)\tilde{\gamma}\cdot\tilde{A}(n)\Psi_n(x)+m_0\bar{\Psi}_n(x)\Psi_n(x)\big],
\end{eqnarray}
where $n$ denotes for  $n_1,\cdots n_p$;  $\partial\!\!\!/$ and $x$ are now used only for coordinates $x_{p+1},\cdots x_{d}$; $\tilde{\gamma}_\mu$ is the gamma matrix associated with coordinates $x_1,\cdots x_p$. In the derivation, the $p$ dimensional $\partial$ left acting on the Hermite function basis $\vec{f}$ transforms into $i \hat{T}$. With the integral $\int d^px$ and the  orthogonal normalization condition of Hermite function, the operator becomes $\hat{P}\hat{T}\hat{P}^\dagger=\Omega_n$. The action in \Eq{eq:fermionactionnew} becomes that of an interacting field theory of the fermion field $\Psi$ in $1+d-p$ dimensions, which interacts with a discretized gauge field in the rest $p$ dimensions, i.e., $\tilde{A}(n)=\Omega_{n_1,\cdots n_p}$. This shows a duality relation between a $1+d$ dimensional free fermion field and a $1+d-p$ dimensional interacting fermion field with interactions coming from the rest $p$ dimensions.

One can also consider the case of $1+d$ dimensional complex scalar field. Similarly, we expand the complex scalar field $\phi$ as follows:
\begin{eqnarray}
\label{eq:expand1}
&&\phi(t,x_1,\cdots x_p,x_{p+1}\cdots x_{d})
=\sum_{n_1,\cdots n_p} \notag\\
&&\Phi_{n_1,\cdots n_p}(t,x_{p+1},\cdots x_{d})[\hat{P}f]_{n_1}(x_1)\cdots[\hat{P}f]_{n_p}(x_p)\,.
\end{eqnarray}
$\Phi$ is a  new field, the part that has not been expanded. Performing the same procedure as in the fermion case above, we obtain the commutation relation for the  new field $\Phi$ as:
\begin{eqnarray}
\label{eq:quanta1}
&&\left[ {\Phi}_{n_1,\cdots n_p}(t,x_{p+1},\cdots x_{d}),\dot{ \Phi}^\ast_{m_1,\cdots m_p}(t,y_{p+1},\cdots y_{d})\right]\notag\\
=&&i\delta^{n_1,m_1}\cdots \delta^{n_p,m_p}\delta(x_{p+1}-y_{p+1})\cdots\delta(x_d-y_d)\,,
\end{eqnarray}
with $\dot{\Phi}$ denoting the time derivative of $\Phi$. It is easy to check that the original field satisfies: $\left[ {\phi}(t,x),\dot{\phi}^\ast(t,y)\right]=i\delta^{d}(x-y)$, the commutation relation for a free complex scalar field can therefore be reproduced. \Eq{eq:quanta1} suggests that the new field $\Phi$ is also a complex scalar field.

Let's move on to consider the action of the $d+1$ dimensional free complex scalar field:
\begin{eqnarray}
\label{eq:scalaraction}
\mathcal{S}=\int dt d^{d}x\left[\partial_\mu\phi^\ast(x)\partial^\mu\phi(x)+m_0^2\phi^\ast(x)\phi(x)\right]\,.
\end{eqnarray}
By inserting \Eq{eq:expand1} into the \Eq{eq:scalaraction}, the action becomes $1+d-p$ dimensional as:
\begin{eqnarray}
\mathcal{S}=&&\int dt d^{d-p}x\sum_n\big[\partial_\mu\Phi_n^\ast(x)\partial^\mu\Phi_n(x)\notag\\
&&+M_n^2\Phi_n^\ast(x)\Phi_n(x)+m_0^2\Phi_n^\ast(x)\Phi_n(x)],
\end{eqnarray}
again, $n$ denotes for  $n_1,\cdots n_p$;  $\partial$ and $x$ are now used only for coordinates $x_{p+1},\cdots x_{d}$. The $1+d$ dimensional scalar field then becomes a $1+d-p$ dimensional scalar field interacting with a tower of massive states with  $M_n^2=\Omega^2_{n_1}+\cdots+\Omega^2_{n_p}$.

\section{Yang-Mills field on Hermite function basis and Matrix model}\label{sec:YM}

When applying the Hermite function basis for Yang-Mills field, it eventually gives an action of  the BFSS matrix model. Firstly, we write here  the Yang-Mills action as~\cite{Alkofer:2000wg}:
\begin{eqnarray}
\label{eq:YM}
\mathcal{S}=-\frac{1}{4}\int d^4x  F^{a}_{\mu\nu} (x)F^{a,\mu\nu}(x),
\end{eqnarray}
where the field strength tensor $F^a_{\mu\nu}$ defined as:
$F^{a}_{\mu\nu} = \partial_{\mu}^{} A^{a}_{\nu} - \partial_{\nu}^{} A^{a}_{\mu}
- gf^{abc} A^{b}_{\mu} A^{c}_{\nu}$,
with $g$ the coupling constant; $A^{a}_{\mu}$ the bosonic gauge field; $f^{abc}$ the structure constant of the gauge group.
Now we can expand $A^a_\mu(x)$ for all three spatial coordinates as:
\begin{eqnarray}
\label{eq:expandA}
A^a\left(t,\vec{x}\right)
&=&  A^{a,n}(t)[\hat{P}f]_{n_1}(x_1)[\hat{P}f]_{n_2}(x_2)[\hat{P}f]_{n_3}(x_3),\notag\\
X_j^a\left(t,\vec{x}\right)
&=&   X^{a,n}_j(t)[\hat{P}f]_{n_1}(x_1)[\hat{P}f]_{n_2}(x_2)[\hat{P}f]_{n_3}(x_3),\notag\\
\end{eqnarray}
with $n=(n_1,n_2,n_3)$, $A^a=A^a_0$ and  $X_j^a=A^a_{j=1,2,3}$. Now inserting \Eq{eq:expandA} into  \Eq{eq:YM}, the action in \Eq{eq:YM} becomes:
\begin{equation}
\label{eq:YM1}
	\mathcal{S}=\mathcal{S}_0+\mathcal{S}_{int}+\mathcal{S}_{im}\,,
\end{equation}
where
\begin{subequations}
\begin{eqnarray}
&&\mathcal{S}_0\notag\\
=&&-\frac{1}{2} \int dt\big{(}\left|\dot{X}_j^{a,n}\right|^2+\Omega_{n_j}^2\left|A^{a,n}\right|^2\notag\\
&&+\frac{1}{2}\big{|}\Omega_{n_{j^\prime}}X^{a,n}_j-\Omega_{n_j}X^{a,n}_{j^\prime}\big{|}^2\big{)},\\
&&\mathcal{S}_{int}\notag\\
=&&-\frac{1}{2}\int dt  \big{(}-2g\tau^{nml}f^{abc}\dot{X}_j^{a,n}  X_j^{c,m}A^{b,l}\notag\\
&&+g^2f^{abc}f^{ab^\prime c^\prime}\lambda^{nmlk}A^{b,n}X^{c,m}_{j}A^{b^\prime,l}X^{c^\prime,k}_{j}\notag\\
&&+\frac{1}{2}g^2f^{abc}f^{ab^\prime c^\prime}\lambda^{nmlk}X^{b,n}_jX^{c,m}_{j^\prime}X^{b^\prime,l}_jX^{c^\prime,k}_{j^\prime}\big{)},\\
&&\mathcal{S}_{im}\notag\\
=&&i\int dt \big{[}\Omega_{n_j}A^{a,n} (\dot{X}^{a,n,\ast}_j-gf^{abc}A^{b,m}X_j^{c,l}\tau^{nml}) \notag\\
&&+gf^{abc}\tau^{nml}(\Omega_{n_{j^\prime}}X^{a,n}_j-\Omega_{n_j}X^{a,n}_{j^\prime})X^{b,m}_{j^\prime}X_{j}^{c,l}\big{]},
\end{eqnarray}
\end{subequations}
with summation of indices $a,b,c$, $j^\prime, j$ and $n,m,l,k$ the order of Hermite function,  and also
\begin{eqnarray}
&&\tau^{nml}\notag\\
=&&\prod_{i=1,2,3}\int d^3x  [\hat{P}f]_{n_i}(x_i) [\hat{P}f]_{m_i}(x_i)[\hat{P}f]_{l_i}(x_i),\\
&&\lambda^{nmlk}\notag\\
=&&\prod_{i=1,2,3}\int d^3x [\hat{P}f]_{n_i}(x_i)[\hat{P}f]_{m_i}(x_i)[\hat{P}f]_{l_i}(x_i)[\hat{P}f]_{k_i}(x_i).\notag
\end{eqnarray}
One can see that the procedure defines an exact way of dimensional reduction which leads directly from the 3+1 dimensional Yang Mills theory to a one dimensional theory, \Eq{eq:YM1} .

Specifically, the free field with $g=0$ becomes a tower of massive gauge fields, similarly as the cases in the  previous section. One can now have an adventurous  guess that for the strong coupling limit, the strongly coupled field can be described only by Hermite function of the lowest order  $n_i,m_i,l_i,k_i=0$, and then $\Omega_{n_j}=0$ correspondingly. With this assumption, one can greatly reduce the action as:
\begin{eqnarray}
\label{eq:YMre}
\mathcal{S}=&&-\frac{1}{2}\int dt\biggl{(}\left( \dot{X}_j^{a}\right)^2-\frac{2g\omega^{3/2}}{\left(\frac{9}{4}\pi\right)^{3/4}}f^{abc}\dot{X}_j^{a}  X_j^{c}A^{b}\notag\\
&&+\frac{g^2\omega^{3}}{\left(2\pi\right)^{3/2}}\left[A,X_j\right]^2+\frac{g^2\omega^{3}}{2(2\pi)^{3/2}}g^2\left[X_{j^\prime},X_j\right]^2\biggl{)},
\end{eqnarray}
with $X_j^a, A^b$  standing for the lowest order  of the gauge field,  and $\left[X_i,X_j\right]=f^{abc}X^b_iX^c_j$. Now if setting $g^\prime=g/\left(\frac{9}{4}\pi\right)^{3/4}$ with rescaling the fields as $(X_j^a,A^b)\rightarrow \omega^{-1/2}( X_j^{a},A^b)$, and $DX^a_j=\dot{X}^a_j-g^\prime\omega\left[A,X_j\right]$, one can rewrite the action as:
\begin{eqnarray}
\label{eq:YMre1}
\mathcal{S}=&&-\frac{1}{2}\int dt\biggl{(}\frac{\left(D{X_j}^{a}\right)^2}{\omega}+\frac{\omega}{2}\left(9/8\right)^{3/2}g^{\prime}{}^2\left[X_{j^\prime},X_j\right]^2\notag\\
&&+\left[\left(9/8\right)^{3/2}-1\right]\omega g^{\prime}{}^2\left[A,X_j\right]^2\biggl{)},
\end{eqnarray}
the first line is just the bosonic part of the BFSS matrix model, and the second term is an additional correction. Note that the higher order correction coming from the higher orders of Hermite functions  converges fast since the prefactor of  coupling  becomes smaller. This approach thus offers an systematic way of computing quantities of strongly coupled Yang-Mills theory with the perturbative series now also involving  the order of Hermite function.

 %With comparison to the Matrix model from string theory, it is interesting that one can determine the slope parameter:
%\begin{equation}
%\alpha^\prime=\frac{1}{2\pi}(\frac{8}{9})^{\frac{3}{4}}
%\end{equation}

\section{summary}\label{sec:summary}

We depict a novel duality phenomenon through the Hermite function basis.
   The main idea is to apply the Hermite function basis instead of the plane wave function basis for the expansion. By doing so, the free propagator becomes a tower of propagators with additional mass terms.
 Since the Hermite function naturally defines the basis of the interacting fields, we construct an equivalent quantization procedure on this basis.   In detail, we find that the action of a $1+d$ dimensional free fermionic field is dual to the action of a tower of $1+d-p$ dimensional fermionic fields $\Psi_n$ coupled to the constant gauge field $\tilde{A}(n)$ in the rest $p$ dimensions. Similarly, for the scalar field, a $1+d$ dimensional free scalar field is dual to a tower of $1+d-p$ dimensional massive scalar fields with additional mass terms coming from the $p$ dimensions.  In the sense of   duality, here one theory is achieved by expanding the other theory on the Hermite function basis.

Now, although the duality considered here looks very similar to the particle-vortex duality, the direct relation has not yet been built. Heuristically, the attached flux in the particle vortex duality is related to the  interacting field, which coincides with the interpretation of the monopole operator, and hence, the duality is possibly achieved by  expanding the gauge field on the Hermite function basis.  On the Hermite function basis,  the original gauge field can be interpreted as the string-like interaction. Besides,  such an interaction  is presumably anchored  through the topological term which is usually required in the particle-vortex duality.   This needs to be further investigated in the future.

Nevertheless,  this approach can be broadly applied to the study of duality phenomena. Here we apply it in particular to the 3+1 dimensional Yang Mills theory. After applying the Hermite function to reduce the three spatial dimensions, an exact form of the corresponding one-dimensional theory can be obtained. If the Hermite function of the lowest order is applied, the resulting action is found to become the BFSS matrix model.
 %Furthermore, with comparison to the matrix model, one can then determine the string slope parameter as $\alpha^\prime=\frac{1}{2\pi}(\frac{8}{9})^{\frac{3}{4}}$.
  Here we still leave out the topological term, which will be considered in further studies.

On the practical side, the Hermite function basis provides a possible way of approaching Minkowski space non-perturbatively. Non-perturbative calculations are usually performed in Euclidean space, where the  information in Minkowski space, and in particular the analytic properties of the non-perturbative propagators on plane wave basis, becomes very complicated. The Hermite function basis shows the ability to deal with different kinds of propagators, and therefore through the basis, the information in Minkowski space can potentially be accessed directly. The momentum representation is also readily accessible as the Hermite function is the eigenfunction of the Fourier transform.

\section{Acknowledgement}

M. Ding is grateful for support by Helmholtz-Zentrum Dresden-Rossendorf High Potential Programme. Y. Liu is supported by the National Natural Science Foundation of China under Grant Nos. 11175004 and 12175007.

\bibliography{ref-lib}
\end{document}